\documentclass[reprint,superscriptaddress,10pt, amsmath,amssymb, aps]{revtex4-2}
\pdfoutput=1 
\usepackage[utf8]{inputenc}
\usepackage[T1]{fontenc}
\usepackage{graphicx}             
\usepackage{xcolor}
\definecolor{darkblue}{rgb}{0, 0, 0.8}
\usepackage[colorlinks=true, breaklinks=true, linkcolor=darkblue, citecolor=darkblue, urlcolor=darkblue]{hyperref}
\usepackage{braket}
\usepackage{soul} 

\newcommand{\beq}{\begin{equation}}
\newcommand{\eeq}{\end{equation}}
\newcommand{\beqa}{\begin{eqnarray}}
\newcommand{\eeqa}{\end{eqnarray}}

\newcommand{\um}{\mu\textrm{m}}
\newcommand{\vtg}{V_\mathrm{{TG}}}
\newcommand{\vbg}{V_\mathrm{{BG}}}
\newcommand{\vL}{V_\mathrm{{L}}}
\newcommand{\vR}{V_\mathrm{{R}}}
\newcommand{\bfA}{{\textbf{a}}}
\newcommand{\bfB}{{\textbf{b}}}
\newcommand{\bfC}{{\textbf{c}}}
\newcommand{\bfD}{{\textbf{d}}}
\newcommand{\bfE}{{\textbf{e}}}
\newcommand{\bfF}{{\textbf{f}}}
\newcommand{\bfG}{{\textbf{g}}}
\newcommand{\bfH}{{\textbf{h}}}
\newcommand{\bfI}{{\textbf{i}}}

\newcommand{\beginsupplement}{%
        \setcounter{table}{0}
        \renewcommand{\thetable}{S\arabic{table}}%
        \setcounter{figure}{0}
        \renewcommand{\thefigure}{S\arabic{figure}}%
     }
     
\usepackage{enumitem}
\setlist[description]{leftmargin=*}

\usepackage{hyperref}
\hypersetup{colorlinks=true,linkcolor=blue,citecolor=blue,filecolor=green,urlcolor=blue}

\let\oldparagraph\paragraph
\renewcommand{\paragraph}[1]{\oldparagraph{\textbf{#1}}}

\begin{document}

\title{Quantum control of exciton wavefunctions in 2D semiconductors}

\author{Jenny Hu}
\thanks{E.L. and J.H. contributed equally}
\affiliation{Department of Applied Physics, Stanford University, Stanford, CA 94305, USA}
\affiliation{SLAC National Accelerator Laboratory, Menlo Park, CA 94025, USA}

\author{Etienne Lorchat}
\thanks{E.L. and J.H. contributed equally}
\affiliation{NTT Research, Inc. Physics \& Informatics Laboratories, 940 Stewart Dr, Sunnyvale, CA 94085}

\author{Xueqi Chen}
\affiliation{Department of Physics, Stanford University, Stanford, CA 94305, USA}
\affiliation{SLAC National Accelerator Laboratory, Menlo Park, CA 94025, USA}

\author{Kenji Watanabe}
\affiliation{Research Center for Functional Materials, National Institute for Materials Science, 1-1 Namiki, Tsukuba 305-0044, Japan}

\author{Takashi Taniguchi}
\affiliation{ 
International Center for Materials Nanoarchitectonics, National Institute for Materials Science, 1-1 Namiki, Tsukuba 305-0044, Japan}

\author{Tony F. Heinz}
\affiliation{Department of Applied Physics, Stanford University, Stanford, CA 94305, USA}
\affiliation{SLAC National Accelerator Laboratory, Menlo Park, CA 94025, USA}

\author{Puneet A. Murthy}
\email{murthyp@ethz.ch}
\affiliation{NTT Research, Inc. Physics \& Informatics Laboratories, 940 Stewart Dr, Sunnyvale, CA 94085}
\affiliation{Institute for Quantum Electronics, ETH Z\"urich, CH-8093 Z\"urich, Switzerland}

\author{Thibault Chervy}
\email{thibault.chervy@ntt-research.com}
\affiliation{NTT Research, Inc. Physics \& Informatics Laboratories, 940 Stewart Dr, Sunnyvale, CA 94085}

\date{August 11, 2023}

\maketitle

\textbf{Excitons --bound electron-hole pairs-- play a central role in light-matter interaction phenomena, and are crucial for wide-ranging applications from light harvesting and generation to quantum information processing. A long-standing challenge in solid-state optics has been to achieve precise and scalable control over the quantum mechanical state of excitons in semiconductor heterostructures. Here, we demonstrate a technique for creating tailored and tunable potential landscapes for optically active excitons in 2D semiconductors that enables in-situ wavefunction shaping at the nanoscopic lengthscale. Using nanostructured gate electrodes, we create localized electrostatic traps for excitons in diverse geometries such as quantum dots and rings, and arrays thereof. We show independent spectral tuning of multiple spatially separated quantum dots, which allows us to bring them to degeneracy despite material disorder. Owing to the strong light-matter coupling of excitons in 2D semiconductors, we observe unambiguous signatures of confined exciton wavefunctions in optical reflection and photoluminescence measurements. Our work introduces a new approach to engineering exciton dynamics and interactions at the nanometer scale, with implications for novel optoelectronic devices, topological photonics, and many-body quantum nonlinear optics.}

\subsection*{Introduction}

\begin{figure}[htbp]
\begin{center}
\includegraphics[width=\linewidth]{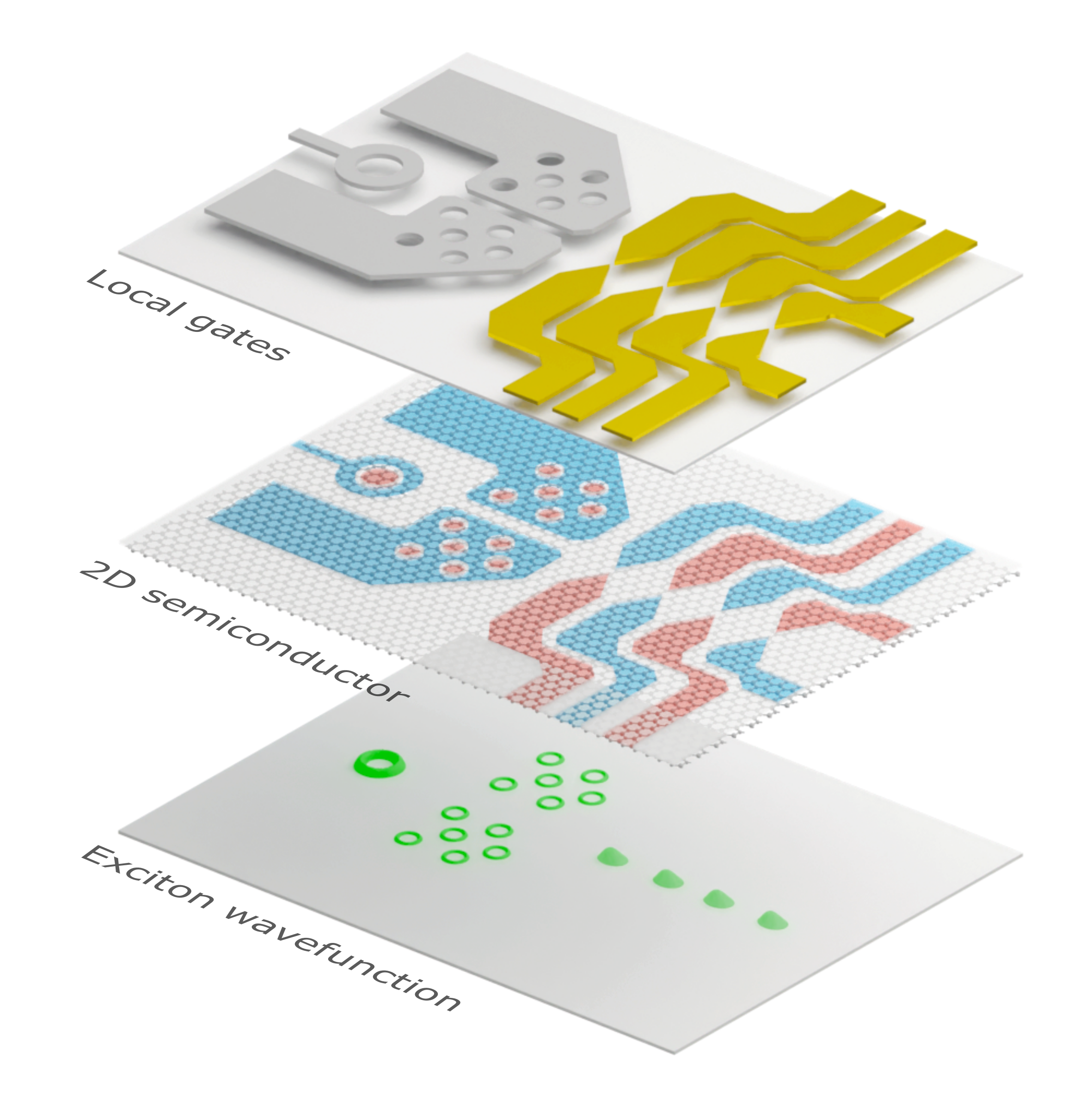}
\caption{\textbf{Quantum excitonics in 2D semiconductors}. Schematic illustration of our approach. \textit{Upper layer:} Nanostructured gate electrodes are positioned in the vicinity ($10${\it s} of nm) of a 2D semiconductor. \textit{Middle layer} Voltages applied to the gates define lateral distributions of in-plane electric fields $\vec{F}(\mathbf{r})$ and charge densities $\rho(\mathbf{r})$ for itinerant holes (red) and electrons (blue) in the 2D semiconductor. \textit{Lower layer} Exciton confinement occurs due to a combination of electric field-induced Stark shift and exciton-charge interactions (see Eq.\,\ref{eq:TotalPotential}). This enables in-situ control of the center-of-mass exciton wavefunction in arbitrary traps such as quantum rings, quantum dots and scalable arrays.}
\label{fig:one}
\end{center}
\end{figure}

The ability to trap and manipulate the quantum mechanical state of particles is a central pillar across various disciplines of quantum science, including ultracold atoms \cite{Gross2017, Kaufman2021}, ion traps \cite{Bruzewicz2019} and superconducting qubits \cite{Kjaergaard2020}. In semiconductor optics, the traditional approach to confinement of optical excitations involves nanostructures such as self-assembled or colloidal quantum dots, which are typically fabricated through growth or implantation techniques. These methods lead to ensembles of quantum dots (or rings) with random positions and broad energy distribution with limited local control \cite{doi:10.1126/science.aaz8541, Aharonovich2016, Azzam2021}. The resulting lack of scalability has hindered their widespread technological implementation. In addition to these methods, alternative material modulation approaches have recently been explored, such as strain engineering \cite{Palacios2017, Rosenberger2019, Wang2021, Yu2021, Lenferink2022}, electron and ion beam irradiation \cite{Fournier2021, Gerard2023}, and moir\'{e} potential engineering \cite{Seyler2019, Bai2020, Zhang2021, Susarla2022}, which offer varying degrees of control on the positions, energies or number of emitters. Furthermore, electrostatic confinement of spatially indirect excitons in semiconductor heterostructures has been reported \cite{High2009, Shanks2021}; however, their vanishing coupling to light limits potential photonics applications. The ability to precisely control the individual quantum states of optically active excitons with strong light-matter coupling has thus remained a major experimental challenge. 

Recently, a novel technique has been reported for confining direct excitons purely using electric fields and charge density gradients in 2D semiconductor heterostructures \cite{Thureja2022}. The method relies on the fact that in-plane electric fields ($\vec{F}(\mathbf{r})$) induce a quadratic dc Stark shift of the excitons, and charge density ($\rho(\mathbf{r})$) leads to an interaction-induced density-dependent energy shift, according to 
\begin{equation}
    \Delta E = -\frac{1}{2}\alpha |\vec{F}(\mathbf{r})|^2 + \beta\rho(\mathbf{r}),
    \label{eq:TotalPotential}
\end{equation}
where $\alpha$ is the dc polarizability of excitons and $\beta$ is the exciton-electron coupling constant. Spatial variations of doping densities and electric fields can thus be used to trap excitons. Based on this effect, quantum confinement was demonstrated in a gate-defined lateral p-i-n junction. However, as the trapping occurs along the gate edges, the technique has so far been limited to excitonic 1D quantum wires. 

Here, we demonstrate scalable and tunable electrostatic traps of arbitrary shapes for direct excitons in monolayer Transition Metal Dichalcogenide (TMD) semiconductors. We illustrate our approach in Fig.\,\ref{fig:one}. The crux of our work lies in using lithographically nano-structured gate electrodes in proximity to the 2D semiconductor plane (upper layer). This allows to precisely define the lateral distribution of in-plane electric fields $\vec{F}(\mathbf{r})$ and density $\rho(\mathbf{r})$ of itinerant holes (red) and electrons (blue) in the 2D semiconductor, with a resolution down to a few tens of nanometers (middle layer). This in turn enables tailor-made landscapes for excitons with a high degree of control on the wavefunction profile (lower layer). To highlight the versatility of our technique, we focus on two exemplary trap geometries, namely quantum rings and quantum dots, that may serve as building blocks for more extended landscapes. Furthermore, we demonstrate the unprecedented scalability of our approach by realizing arrays of quantum rings and independently controlled quantum dots which is of particular relevance for future optoelectronics and photonics technologies.


The basic structure of our exciton trapping devices consists of a monolayer TMD semiconductor, such as MoSe$_2$, encapsulated by hexagonal boron nitride (hBN) spacers of appropriate thickness. This heterostructure is stacked on a Si/SiO$_2$ substrate with top (TG) and bottom (BG) gate electrodes, which can either be graphene (Gr) or metallic thin films. We nanostructure one of the gate electrodes through a combination of electron beam lithography and dry etching, which allows for patterning resolutions of around 50 nm. A global BG is used to define the charge configuration of the entire monolayer, whereas the nanostructured TGs enable local control of charge densities and fields. Even though either of the gates can be patterned without losing functionality, we choose to define the structures on the TGs and keep the BG for global doping, for ease of fabrication. Our devices are cooled down to $\sim5\,$K in a closed cycle dry cryostat with optical access. To characterize the excitonic states in our system, we mainly rely on optical broadband reflection and photoluminescence spectroscopy with a spatial resolution of $0.7\,\um$. A detailed account of our fabrication and experimental procedures is described in the Supplementary Information (SI \ref{Fabrication method},\ref{Experimental setup}).

\begin{figure*}[htbp]
\begin{center}
\includegraphics[width=\linewidth]{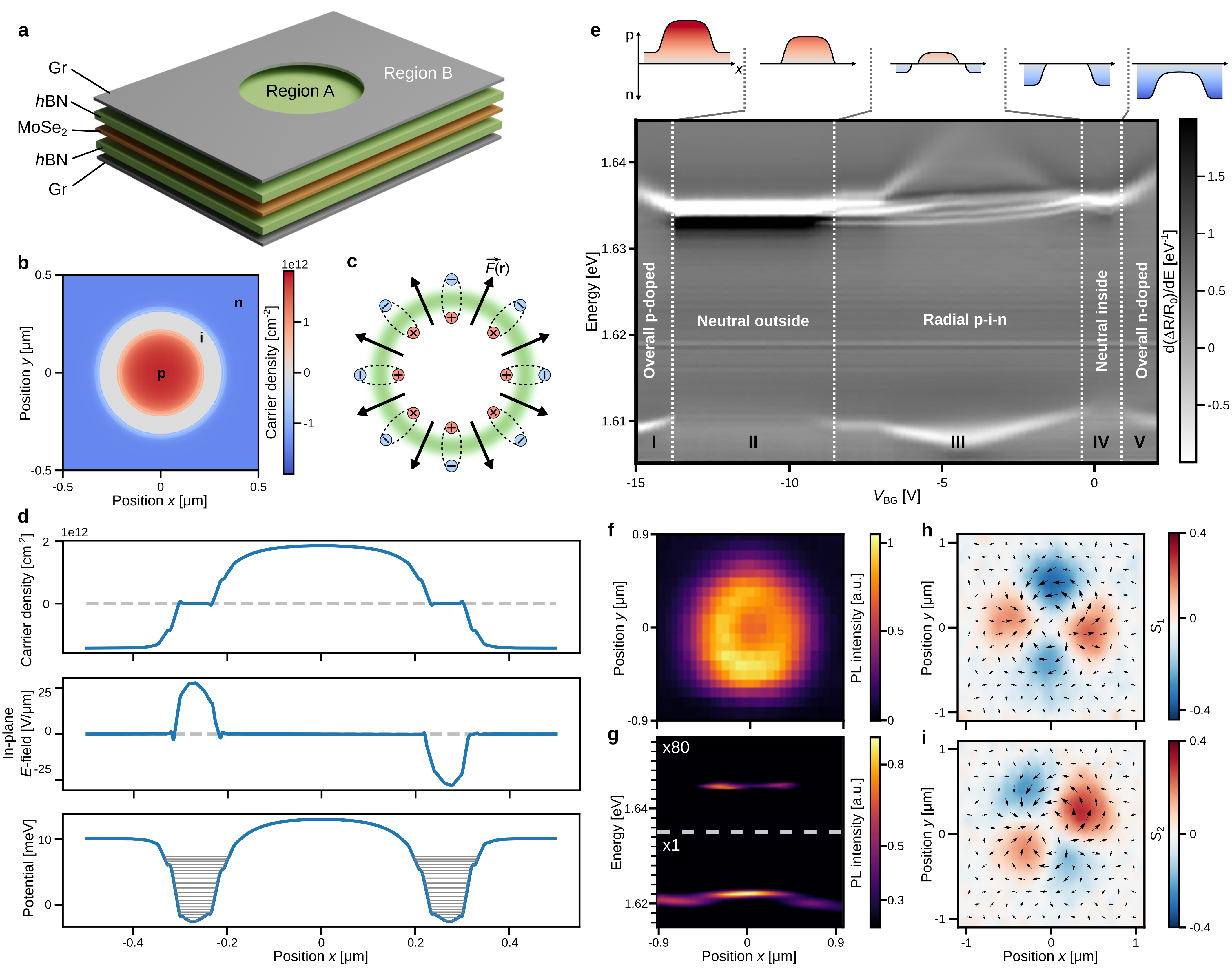}
\caption{\textbf{Ring traps for excitons}. 
(\textbf{a}) Dual-gated TMD heterostructure where a nano-hole etched in the top gate defines quantum ring confinement for excitons. 
(\textbf{b}) Finite element electrostatic simulations of the device for a $600\,$nm gate hole with gate voltage configuration  $(\vbg,\vtg) = (-5.0\,\mathrm{V},9.5\,\mathrm{V})$; Charge distribution shows a globally electron doped semiconductor with holes trapped in the center of the nano-hole, surrounded by a ring-shaped neutral region. 
(\textbf{c}) Schematic showing radially dipolar excitons quantum confined in a ring. 
(\textbf{d}) Finite element simulations of the charge distribution (\emph{upper panel}), field distribution (\emph{middle panel}) and exciton trapping potential (\emph{lower panel}) for a 600 nm diameter hole device. The horizontal gray lines in the lower panel are the c.o.m eigenstates of the 2D Schr\"{o}dinger equation. Gate voltage configuration is the same as in (\textbf{b}).
(\textbf{e}) First-derivative reflectance contrast spectra as a function of $\vbg$, for fixed $\vtg = 9.5\,$V. Charging configurations for each regime (I - V) are illustrated above. 
(\textbf{f}) Spatially resolved scan of PL emission integrated over the confined states acquired on a 1 $\mu$m diameter hole. 
(\textbf{g}) Spectral cross cut through the center of the hole ($y=0$), showing the trion emission from regions A and B and neutral exciton emission of ring states. (\textbf{h,i}) Stokes vector analysis of the confined state emission map in the linear polarization basis, which demonstrates that PL emission of ring states is polarized in the radial-azimuthal basis. Arrows represent the local linear Stokes vector $\textbf{S} = (S_1,S_2)$.}
\label{fig:two}
\end{center}
\end{figure*}

\subsection*{Excitonic quantum rings}
In Fig.\,\ref{fig:two}, we describe our scheme to create ring traps for excitons. An annular confinement geometry can be obtained simply by patterning circular holes in the TG, as shown schematically in Fig.\,\ref{fig:two}\,\bfA. In order to obtain the strongest confinement, we apply opposite voltages on the top and bottom gates to reach the annular p-i-n regime, where a central p-doped puddle is enclosed in a Fermi sea of electrons with a ring-shaped neutral region separating them, as seen in Fig.\,\ref{fig:two}\,\bfB. The combination of in-plane fields and charge gradients leads to tight confinement of excitons in this annular region (\ref{eq:TotalPotential}). Furthermore, the in-plane fields lead to radially polarized in-plane dipolar excitons as illustrated in Fig.\,\ref{fig:two}\,\bfC.

In Fig.\,\ref{fig:two}\,\bfD, we show results of finite element simulations of the charge and field distributions in the annular p-i-n configuration for a top gate hole with diameter $600\,$nm, which provides a quantitative understanding of the electrostatics of the device. We calculate the confinement potential according to Eq.\,\ref{eq:TotalPotential}, which shows the expected Mexican hat-like profile. We note that the dominant contribution to the potential comes from the exciton-electron interaction (second term in Eq.\,\ref{eq:TotalPotential}) whereas the Stark shift has a relatively mild effect. Solving the 2D Schr\"{o}dinger equation in the center-of-mass (c.o.m) frame of excitons, we obtain discrete radial trap levels, with a level separation of $\sim 0.3-0.8\,$meV. We present a detailed account of the trap profile as a function of voltage in the SI (section \ref{Simulation600nmhole}).

In order to verify the nature of excitonic states in this system, we now move on to optical measurements of the device. In monolayer MoSe$_2$, the onset of electron or hole doping is associated with the emergence of a bound trion state and a density-dependent blue shifting exciton state (repulsive polaron) \cite{Sidler2017, Efimkin2017, Glazov2020}. Therefore, measuring reflectance spectra as a function of $\vtg$ and $\vbg$ allows to precisely distinguish doping configurations inside and outside the TG hole. Region A (inside the hole) is primarily affected by the BG, whereas region B (outside the hole) can be doped using both TG and BG. 

In Fig.\,\ref{fig:two}\,\textbf{e}, we present the measured reflectance spectra obtained from a $600\,$nm diameter hole as a function of $\vbg$, for a fixed $\vtg = 9.5\,$V. In order to highlight the faint signatures of quantum confined states, the reflectance contrast $\Delta R/R_0$ is differentiated with respect to energy as detailed in SI \ref{Differential reflectivity data analysis}. Five distinct regimes can be identified as the applied back gate voltage $\vbg$ is swept. For positive $\vbg$, the action of the back gate reinforces that of the top gate, leading to overall n-doping of the device (regime V in Fig.\,\ref{fig:two}\,\bfE). As expected, we observe a trion branch at $\sim 1.610\,$eV, and a repulsive polaron branch at $\sim 1.635\,$eV. Decreasing $\vbg$ reduces the electron density, until we reach the i-i-n regime where region A is neutral (regime IV), as evidenced by the transfer of oscillator strength from the trion branch to the neutral exciton state. As we decrease $\vbg$ further, we reach the p-i-n regime where region A is p-doped and region B is n-doped (regime III), with an annular neutral region in between. Here, we observe repulsive polaron branches from both A and B regions, associated to the two types of carriers. In addition, we observe narrow and discrete resonances emerging from the neutral exciton state. We attribute these new resonances to 1D quantum ring states of different radial mode orders existing at the periphery of the nanohole. Upon decreasing the bottom gate voltage even further, the confined exciton states merge back with the 2D exciton continuum, as the region B outside the dot reaches charge neutrality (regime II). The energy separation between the discrete states is approximately $0.5\,$meV, in agreement with the simulations, which corresponds to a trapping length of $\sim 10\,$nm.

To investigate the spatial profile of the trap, we now turn to a larger TG hole of diameter $1\,\mu$m, which is large enough to optically resolve the different regions A and B with a confocal microscope setup. Fig.\,\ref{fig:two}\,\bfF ~shows a spatially resolved photoluminescence (PL) map of the hole in the radial p-i-n configuration ($\vbg = -2.5\,$V, $\vtg= +9.0\,$V). The emission is filtered at the energy of the confined states. As expected, we observe a clear doughnut-shaped emission profile. The PL spectra taken along a cross-cut through the center of the hole (Fig.\,\ref{fig:two}\,\bfG) shows distinct trionic emission from regions A and B, and confined states emission from the annular neutral region. 

Further insights into the quantum ring states can be obtained by studying the polarization texture of the emitted photons. Previous studies on 1D confined states have shown that electron-hole exchange interactions, coupled with tight confinement of the excitonic wavefunction along one direction, leads to strong polarization splitting in a linear basis \cite{Thureja2022}. Applying this argument to our circular geometry, the ring confinement should induce polarization splitting in the radial-azimuthal basis. To investigate this behavior, we record the PL emission intensity from the lowest confined exciton as a function of position, for different polarization states (see SI \ref{Experimental setup}). From these measurements, the normalized linear polarization Stokes vectors $\textbf{S} \equiv (S_1, S_2) = (I_x - I_y, I_d - I_{ad})/I_0$ can be reconstructed as a function of position on the ring, where $I_x$, $I_y$, $I_d$ and $I_{ad}$ respectively correspond to PL intensity along $x$, $y$, diagonal ($45^\circ$), and anti-diagonal ($-45^\circ$) directions, and $I_0$ is the total PL intensity. In Fig.\,\ref{fig:two}\,\bfH ~and \bfI, we show $S_1$ and $S_2$ for the lowest confined state which both exhibit two-fold symmetric patterns as expected for an azimuthally polarized emitter. These observations clearly demonstrate the quantum confinement of excitons in ring-shaped potentials, and constitute the first major result of this work. 

\begin{figure*}[htbp]
\begin{center}
\includegraphics[width=\linewidth]{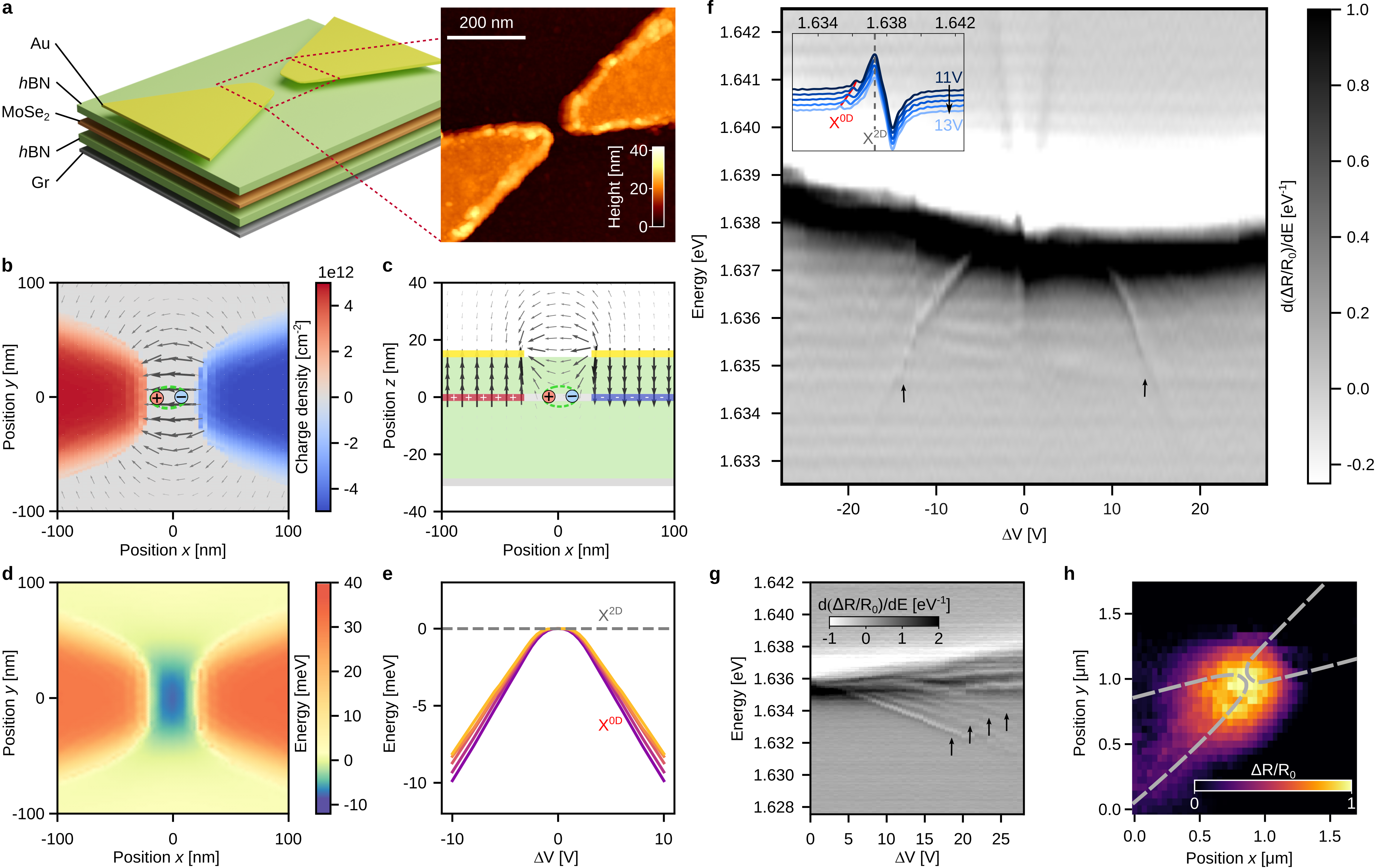}
\caption{\textbf{Bow tie traps for excitons}.
(\textbf{a}) Dual-gated TMD heterostructure with top gate structured as a bow tie. Inset: AFM map of a $35\, $nm bow tie gate. 
(\textbf{b - d}) Electrostatic simulations with $(\vL, \vR, \vbg) \equiv (5.0\,\text{V}, -5.0 \,\text{V}, 0\,\text{V})$. 
(\textbf{b}) Top view ($xy$) of charge density and in-plane electric field at the TMD plane, showing n-doped (blue) and p-doped (red) regions. 
(\textbf{c}) Side view ($xz$) of the electric field at a vertical cut along $y=0$.  
(\textbf{d}) Confining potential for excitons in the gap region. 
(\textbf{e}) Calculated dependence of resonance energy of ladder of states as a function of $\Delta V = \vL - \vR$. 
(\textbf{f}) First-derivative reflectance contrast spectroscopy of the bow tie gap region, as a function of $\Delta V$. The inset shows vertical cross-cuts from $\Delta V = 11\, $V (dark blue) to $\Delta V = 13\, $V (light blue). The red line is a guide to the eye following the confined state. 
(\textbf{g}) First-derivative reflectance contrast of a $100\, $nm gap bow tie, as a function of $\Delta V$, showing a ladder of confined states. 
(\textbf{h}) Spatial dependence of the first-derivative reflectance contrast, integrated over the 0D exciton signal at $\Delta V = 9.9\, $V. Here, $\vbg = 1.5\, $V is applied in order to further energetically separate the 0D state from the X$^{2D}$ signal. This leads to a slight elongation of the states towards the left electrode (SI \ref{BT0Dand1D}).}
\label{fig:three}
\end{center}
\end{figure*}

\subsection*{Tunable quantum dots}
We now turn to the demonstration of tunable quantum dot-like confinement, which has been a major goal in quantum photonics \cite{Chen2014, Trotta2015, Wan2019}. To generate electrostatic 0D nanotraps, we need a gate geometry that allows for tight confinement in both lateral directions. While the gate hole structure described in Fig.\,\ref{fig:two} may also be used for this purpose, the tightness of confinement and energy tunability is significantly limited by fabrication constraints. To achieve tunable 0D traps, we design a bow tie electrode structure as illustrated in Fig.\,\ref{fig:three}\,\bfA, which concentrates electric fields in the nanoscopic gap between the electrodes. These bow ties are fabricated using electron beam lithography and deposition of $10\,$nm Au, with gap sizes ranging from $30-100\,$nm. An atomic force microscope micrograph of a bow tie ($\sim 35\,$ nm gap) is shown in the inset. 

In this geometry, we operate primarily in the neutral regime, where the semiconductor is globally set to neutrality using the bottom gate, and a bias voltage $\Delta V = \vL - \vR $ is applied between the bow tie electrodes to obtain in-plane fields in the nanogap. An electrostatic simulation of the charge and field distributions in the device is shown in Fig.\,\ref{fig:three}\,\bfB ~and \bfC ~(top and side view, respectively; red: holes, blue: electrons). In general, the top hBN thickness should be as small as possible ($\lesssim 20\,$nm) in order to maximize the in-plane component of the electric field on the TMD plane in the gap region.

In Fig.\,\ref{fig:three}\,\bfD, we show the expected 2D trapping potential for excitons in bow ties computed from Eq.\,\ref{eq:TotalPotential}. First, we note that in contrast to the ring traps, the dominant contribution to confinement in the bow tie geometry comes from the electric field-induced Stark shift (first term in Eq.\,\ref{eq:TotalPotential}). Further, the trapping potential features $x-y$ anisotropy, which can be adjusted by varying the width of the electrode tip and the gap size. By solving the 2D Schr\"{o}dinger equation for the exciton c.o.m motion in such a potential, we obtain the expected dependence of the quantum confined states energies on $\Delta V$, shown in Fig.\,\ref{fig:three}\,\bfE. 

To experimentally observe quantum dot states in the bow tie structure, we measure reflectance spectra in the gap region as a function of $\Delta V$. These spectra are acquired from a bow tie with a gap size of $\sim 35\,$nm. Since the gap size is much smaller than the diffraction limit, we expect that any confined states will have a strongly diminished oscillator strength as compared to the 2D exciton resonance. In Fig.\,\ref{fig:three}\,\bfF, we show the first-derivative reflectance contrast spectra $d(\Delta R/R_0)/dE$ as a function of $\Delta V$. For simplicity, we apply symmetric voltages on the two bow tie electrodes; eg. to obtain $\Delta V = 20\,$V, we apply $\vL = -10\,$V and $\vR = +10\,$V. In addition to the broad neutral 2D exciton background (centered around $X^{2D} \sim 1.638\,$eV) and the blue-shifting polaron branches from the gated regions, we observe much narrower resonances emerging below the neutral exciton (linewidth $\Gamma \lesssim 300\,\mu$eV). These states clearly red shift with $\Delta V$, with a similar dependence to Fig.\,\ref{fig:three}\,\bfE. This observed voltage dependence is in excellent quantitative agreement with our simulations (see SI \ref{SimulationBT0D}). Line cuts at different $\Delta V$ are shown in the inset. The energy dependence of these states on the electric field across the bow tie is strong evidence that they occur in the gap region (see SI \ref{SimulationBT0D}, \ref{BT0Dand1D}). A further compelling signature of quantum confinement in such electrical traps is the emergence of a ladder of states corresponding to discrete energy levels in the trap \cite{Thureja2022}. In Fig.\,\ref{fig:three}\,\bfF, such signatures are barely apparent possibly due to the small gap size leading to vanishing oscillator strength of excited states. To confirm this, we perform the same reflectance measurement on a bow tie with a larger gap size ($\sim 100\,$nm), as shown in Fig.\,\ref{fig:three}\,\bfG. Here we do observe a clear ladder of excitonic states emerging below the 2D exciton continuum, which supports our hypothesis.

To further confirm that the observed resonances indeed originate in the gap region, we perform a spatially resolved scan of reflectivity in the vicinity of the bow tie. For this measurement, a small voltage $\vbg = 1.5\, $V is applied in order to further energetically separate the 0D state from the X$^{2D}$ background, which facilitates the analysis (see SI \ref{BT0Dand1D}). The first-derivative reflectance contrast map, integrated over the narrow exciton resonance, is shown in Fig.\,\ref{fig:three}\,\bfH~which demonstrates that this resonance only appears close to the gap region and vanishes as we move away. We have reproduced these observations across different bow ties in the same device, as well as on several different devices (see SI \ref{BTsamples}), which underlines the robustness of our technique. Taken together, our measurements unambiguously demonstrate electrically tunable 0D quantum confinement in bow tie electrode structures, which constitutes the second key result of this work. 

\begin{figure*}[htbp]
\begin{center}
\includegraphics[width=\linewidth]{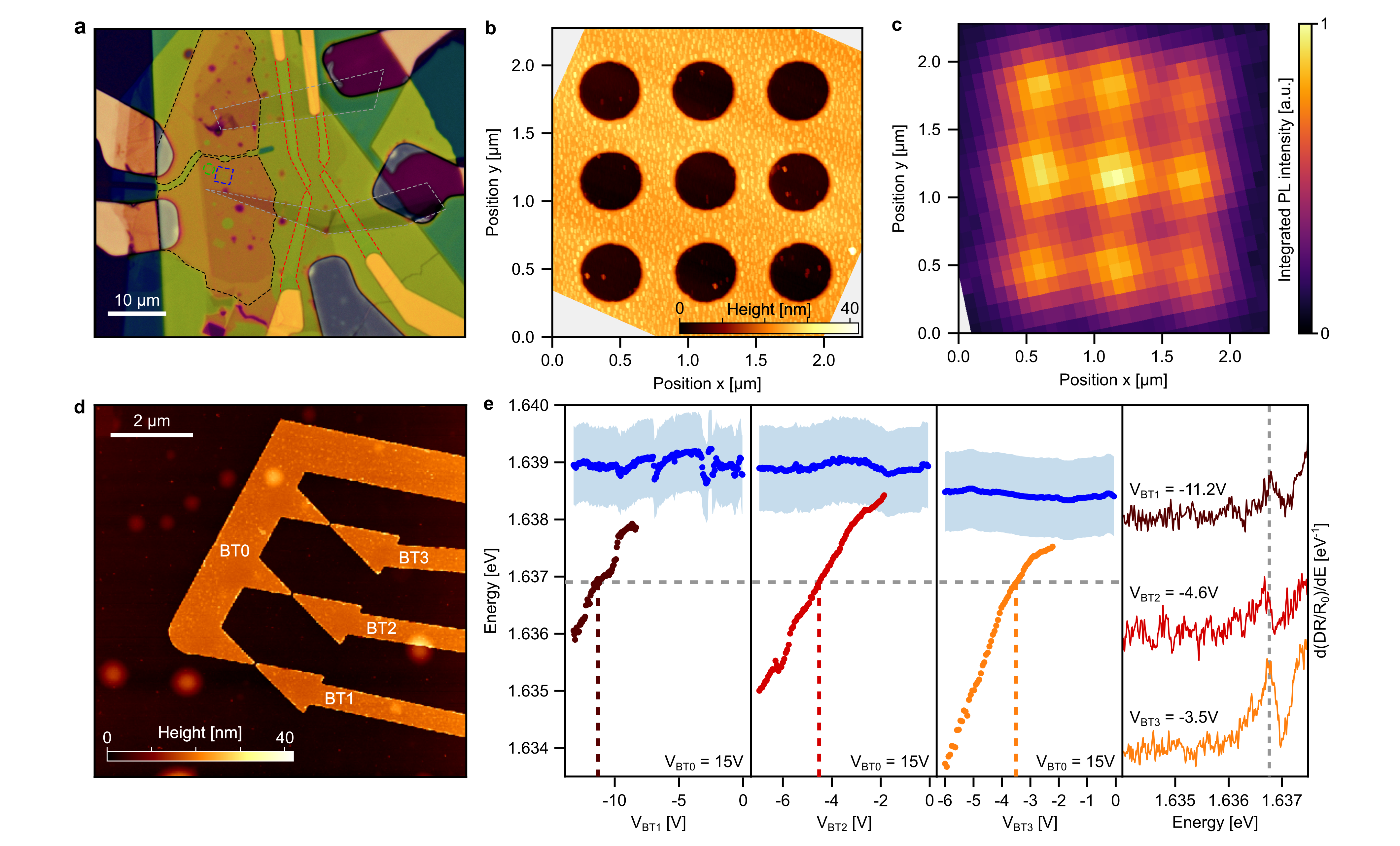}
\caption{\textbf{Scalable arrays}.~ 
(\textbf{a}) Optical micrograph of a van der Waals heterostructure featuring Au bow tie electrodes (red dashed lines) and patterned few layer graphene top gate (black dashed lines). The blue dashed line indicates the $400\, $nm diameter hole array, and the green one highlights the 1 $\mu$m diameter hole studied in Fig.\,\ref{fig:two}. The gray dashed lines indicate the contacts to the TMD. 
(\textbf{b}) AFM topography of the 400 nm hole array. (\textbf{c}) PL map of the hole array, filtered at the confined state emission energy. 
(\textbf{d}) AFM topography of a bow tie array. BT1, BT2 and BT3 are three independent electrode while the counter electrode BT0 is common to the three bow ties. (\textbf{e}) Fitted energies of the 2D exciton (blue dots) and 0D confined states (brown, red, orange dots) for a fixed counter electrode voltage $V_\mathrm{BT0}=15\, $V, as a function of the independent electrode voltages ($V_\mathrm{BT1}$, $V_\mathrm{BT2}$, $V_\mathrm{BT3}$). The shaded blue area indicates the FWHM of X$^{2D}$. {\it Rightmost panel}: Confined states spectra acquired under resonance tuning conditions ($V_\mathrm{BT0}=15\, $V, $V_\mathrm{BT1}=-11.2\, $V, $V_\mathrm{BT2}=-4.6\, $V, $V_\mathrm{BT3}=-3.5\, $V)}
\label{fig:four}
\end{center}
\end{figure*}

\subsection*{Scalable arrays of quantum dots and rings}
The most important advantage of our approach for electrical nanoscale control of exciton wavefunctions is the potential for scaling up to more complex structures. The quantum dot and ring trapping schemes that we presented above can be seen as the building blocks for larger systems. Now, we demonstrate how a variety of nano-structures can be realized in a single TMD heterostructure simply by lithographically defining the desired gate patterns (Fig.\,\ref{fig:four}). Figure\,\ref{fig:four}\,\bfA shows a micrograph of a patterned heterostructure, demonstrating how different gate geometries can be defined on the same device. In Fig.\,\ref{fig:four}\,\bfB, we show an AFM scan of a $3 \times 3$ array of $400\,$nm diameter holes, with a $600\,$nm pitch that allows to optically resolve individual lattice sites. A spatially resolved scan of the PL emission, integrated over the quantum ring state resonances, is shown in Fig.\,\ref{fig:four}\,\bfC. We clearly observe the emission profile of the ring array that matches excellently with the gate pattern. Since the hole diameters are smaller than the diffraction limit, we do not resolve the ring shape of the emission patterns (see Fig.\,\ref{fig:two}\,\bfF for comparison). This shows that ring traps can be scaled up arbitrarily in a single monolayer. 

Next, we demonstrate the scalability of electrically tunable quantum dots. An important motivation for our work is the realization of multiple quantum dots with identical energies, which is a crucial and basic ingredient for several applications, including photonic quantum information processing and quantum communications \cite{Gisin2007}. This has so far remained a hurdle to achieve since existing material modulation approaches are drastically limited by material disorder and process variation \cite{Becher_2023}. We address this problem by fabricating an array of bow ties with independent control for each. An AFM scan of our structure with three bow ties with approximate gap size of $50\,$nm and a separation of $1\,\mu$m is shown in Fig.\,\ref{fig:four}\,\bfD. We electrically short the left-hand side electrodes of the three bow ties (BT0), while maintaining individual control on each of the right-hand side electrodes (BT1, BT2, BT3). This allows to reduce the number of control gates per quantum dot to only one, thus enhancing scalability without compromising on the control over each site. As shown in SI \ref{BTsamples}, we observe the quantum dot confinement signatures described in Fig.\,\ref{fig:three}\,\bfF~in each of the bow ties. 

In Fig.\,\ref{fig:four}\,\bfE, we report the fitted energy of the 0D states as a function of individual gate voltages $V_\mathrm{BT1}$, $V_\mathrm{BT2}$ and $V_\mathrm{BT3}$, while keeping the common counter electrode at $V_\mathrm{BT0} = 15\,$V. The energy of the 2D exciton $X^{2D}$ is shown with blue dots and the linewidth is indicated by the blue shaded regions. The three bow ties show different dependence with voltage, possibly due to material disorder and variations due to fabrication uncertainties. Nevertheless, the three quantum dots can be simultaneously tuned to degeneracy (horizontal dashed line) by applying the suitable voltages across the bow ties (vertical dashed lines). This is clearly evident in the spectra shown in Fig.\,\ref{fig:four}\,\bfF, taken at $V_\mathrm{BT1} = -11.2\,$V, $V_\mathrm{BT2} = -4.6\,$V, and $V_\mathrm{BT3} = -3.5\,$V respectively, which show three quantum dot states resonantly tuned in energy. The ability to combine position-controlled lithographically defined quantum dots, with simultaneous and independent energy tunability into scalable arrays of quantum sites, constitutes the third main result of this work.

\subsection*{Outlook}
We have demonstrated new fundamental building blocks for quantum excitonics. Our approach enables continuous control of the c.o.m  wavefunctions, from micron sized extended states all the way down to quantum confinement into nanoscale dots and rings. Our approach is intrinsically scalable to independently tunable arrays, which is an important feature for future applications. A key advantage of this method is that arbitrary landscapes for excitons can be defined in a non-intrusive manner while retaining the pristine properties of the active material. Hence it can be extended to different semiconductors, including perovskites \cite{Wang2019} and metal chalcogenide compounds \cite{Carey2017}. Furthermore, improvements in fabrication techniques - in particular in lithography - will enable even smaller trapping length scales and better spatial control. 

Our work opens up several avenues for further research. On the fundamental level, an important question is the nature of the confined excitonic state in the relative coordinates, which may be investigated by applying a combination of strong electric and magnetic fields \cite{Fuhrer2002}. Moreover, a detailed study of lifetime and coherence time will provide insights into exciton dynamics in such potentials. The fact that excitons confined in such nanoscopic traps still retain substantial oscillator strength is a major advantage, as this will enable strong coupling to light in microcavities \cite{Najer2019} with expected Rabi splitting in the meV range. Arrays of such electrically confined quantum dots and rings may be combined with microcavity arrays to realize Bose-Hubbard model of photons \cite{Boulier2020}. The potentially enhanced optical nonlinearity of confined exciton systems may allow to go beyond the mean field regime and explore the strong correlations regime, where exotic phases such as Mott insulators are expected. Along the same lines, trapping radially dipolar excitons in ring potentials may have immediate relevance for effecting artificial gauge fields for optical excitations \cite{Togan2018, Chestnov2021}, which is key to realizing topological effects such as a photonic fractional quantum Hall state. From the technological perspective, these configurable exciton landscapes could be of relevance for development of active photonic metamaterials \cite{Takagahara1992} and novel light sources \cite{Guzelturk2014}. \\

\noindent
\textbf{Acknowledgements.} We are grateful to Patrick Kn\"{u}ppel, Deepankur Thureja and Andrea Bergschneider for fruitful discussions. We thank Deepankur Thureja for assistance with finite element simulations, and Elie Vandoolaeghe for help with the experimental setup. \textbf{Funding:} Support for the Stanford/SLAC researchers was provided primarily by the Q-NEXT Quantum Center, a U.S. Department of Energy (DOE), Office of Science, National Quantum Information Science Research Center, with additional support for device fabrication from the DOE, Office of Science, Office of Basic Energy Sciences (BES), Materials Sciences and Engineering Division and the  Gordon and Betty Moore Foundation’s EPiQS Initiative through grant number GBMF9462. J.H. acknowledges support from NTT Research Fellowship. K.W. and T.T. acknowledge support from the JSPS KAKENHI (Grant Numbers 19H05790, 20H00354 and 21H05233).

\noindent
\textbf{Author contributions}
P.A.M. and T.C. conceptualized the work. J.H., X.C. and E.L. fabricated the devices. E.L., J.H. and T.C. performed the measurements and analysed the data. J.H. performed the simulations. P.A.M and T.C wrote the manuscript. K.W. and T.T. provided the hBN crystals. T.F.H., P.A.M and T.C. supervised the project.

\bibliography{References}

\beginsupplement
\section*{Supplementary Information}

\subsection{Experimental setup}
\label{Experimental setup}

\begin{figure*}[htbp]
\begin{center}
\includegraphics[width=0.8\linewidth]{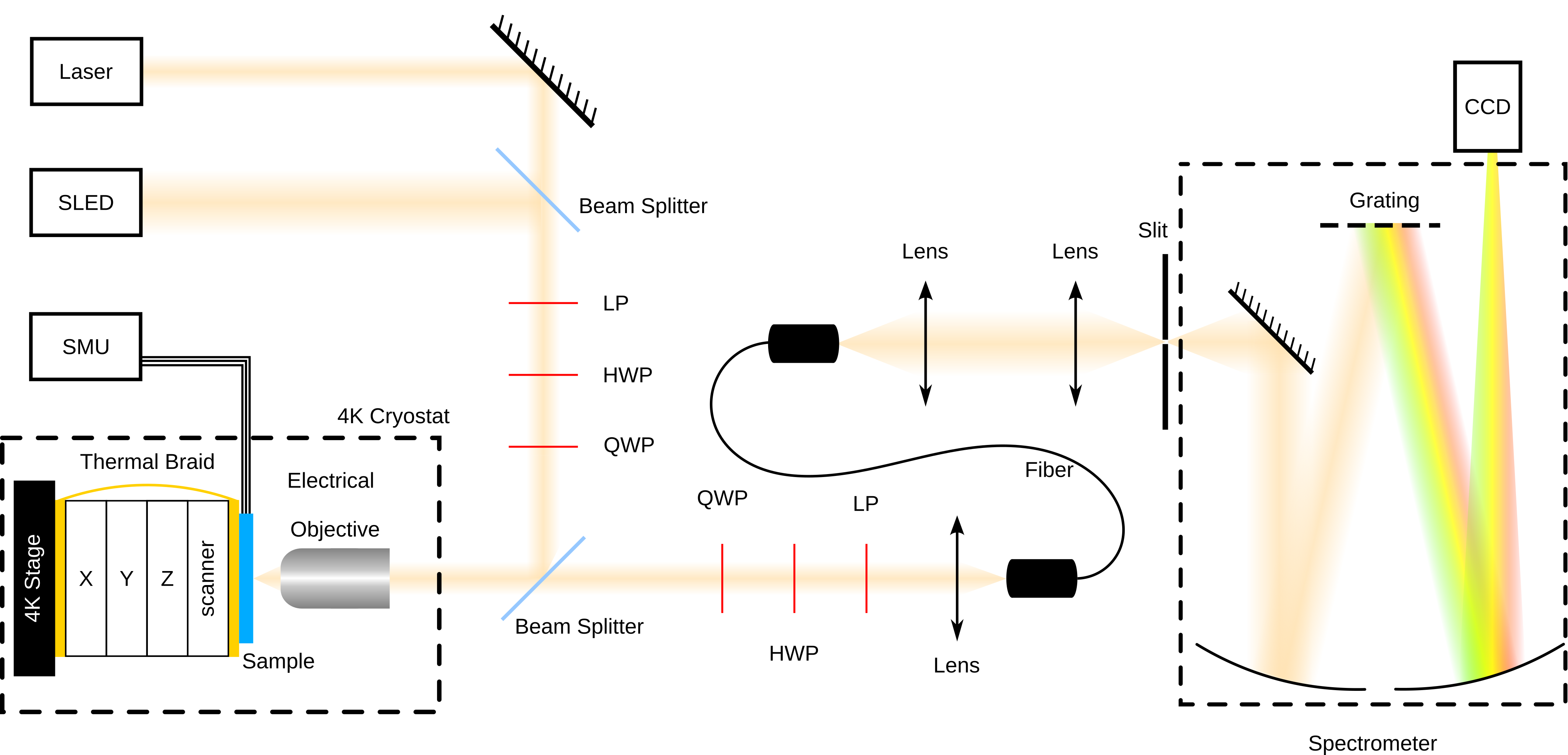}
\caption{Schematic of the optical setup.}
\label{fig:Sone}
\end{center}
\end{figure*}

Figure\,\ref{fig:Sone} depicts our experimental setup. The sample is kept at $\sim5$K using an attoDRY800 closed-loop cryostat from Attocube.
The sample is fixed on a custom made PCB board, which is mounted on a positioning piezo stack composed of three piezo stepper (two Attocube ANPx101/RES/LT and one ANPz102/RES/LT) and a piezo scanner (ANSxyz100std/LT). A cryo-, vacuum-compatible, high numerical aperture (NA = 0.81) objective (Attocube LT-APO/NIR/0.81) is used within the cryostat space. Voltages are applied to the sample gates using a Source Measure Unit (SMU) from Hewlett Packard (HP4142B) fitted with HP 41422B (41420A) modules.

The white light reflectivity measurements are performed using a Superluminescent Light Emitting Diode (SLED Exalos: EXS210065-01). A $720\, $nm Ti:Sa continuous wave tunable laser (Matisse C from Spectra Physics), is used for photoluminescence measurement. The signal beam is collected through a single mode optical fiber (Thorlabs 630HP) that we use as a confocality pin hole. The corresponding collection area on the sample surface has a FWHM of $0.7\, \mu$m which sets the spatial resolution of our optical setup. The fiber output is directed to an imaging spectrometer (Andor SR-750-D1 with SR5-GRT-0600-0500, 600 line per mm grating) equipped with a Peltier cooled CCD camera (Andor DU940P-UV).

Polarization resolved experiments are performed using a set of broadband polarization optics in the excitation and detection arms, both consisting of linear polarizers (Thorlabs LPVIS100-MP2), half wave plates (Thorlabs AHWP10M-980), and quarter wave plates (Thorlabs AQWP10M-980). Linear Stokes vector maps are obtained by recording PL spectra in different linear polarization basis at each point on the sample: 
\begin{align*}
    S_0 = \langle E_x^2 \rangle + \langle E_y^2 \rangle = \langle E_d^2 \rangle + \langle E_a^2 \rangle \\
    S_1 = (\langle E_x^2 \rangle - \langle E_y^2 \rangle)/S_0 \\
    S_2 = (\langle E_d^2 \rangle - \langle E_a^2 \rangle)/S_0 
\end{align*}
where $\langle E_{x(y)}^2 \rangle$ is the intensity of light polarized along $x$ ($y$) direction, and $\langle E_{d(a)}^2 \rangle$ is the intensity of light polarized along the diagonal (antidiagonal) direction. $S_0$ is the total intensity of the signal. 

AFM topographies are obtained in tapping mode, using ScanAsyst-air AFM probes on a MultiMode-8-HR AFM from Bruker.  

\subsection{Differential reflectivity data analysis}
\label{Differential reflectivity data analysis}

Differential reflectivity spectra $\Delta R/R_{0} = (R-R_{0})/R_{0}$ are obtained from the measured reflectivity $R$ by subtracting and dividing by the background spectrum $R_{0}$, measured at the same spot with strong homogeneous doping. This is achieved by keeping the local top gates at 0\,V and applying 10\,V to the global back gate. Where needed, the differential reflectivity spectra are numerically differentiate with respect to energy ($d(\Delta R/R_{0})/dE$) in order to increase the visibility of the confined states resonances.

In Fig.\,\ref{fig:four}\,\bfE, the energy of the 2D exciton and 0D exciton are shown. Those are extracted from fits of $d(\Delta R/R_{0})/dE$ series. The reflectivity spectral lineshape ($S(E)$) of heterostructures as studied here are well approximated by the sum of a pure and dispersive Lorentzian lineshapes ($L_0$ and $L_D$ respectively), defined as follow:

\begin{align*}
    L_0(E) = \frac{A\Gamma}{2\left[ (E - E_0)^2 + (\Gamma/2)^2\right]}\\
    L_D(E) = \frac{A(E - E_0)}{2\left[ (E - E_0)^2 + (\Gamma/2)^2\right]}\\
    S(E) = \cos (\theta)L_0(E) + \sin (\theta)L_D(E)
    \label{eq:CLorentz}
\end{align*}
with $A$ the Lorentzian amplitude, $\Gamma$ the Full Width at Half Maximum (FWHM), $E_0$ the central energy of the mode and $\theta$ the phase between the pure and dispersive Lorentzian. This Lorentzian is then numerically differentiate and fitted to the data, using the least\textunderscore squares routine from the python library scipy.optimize.

\begin{figure*}[htbp]
\begin{center}
\includegraphics[width=0.8\linewidth]{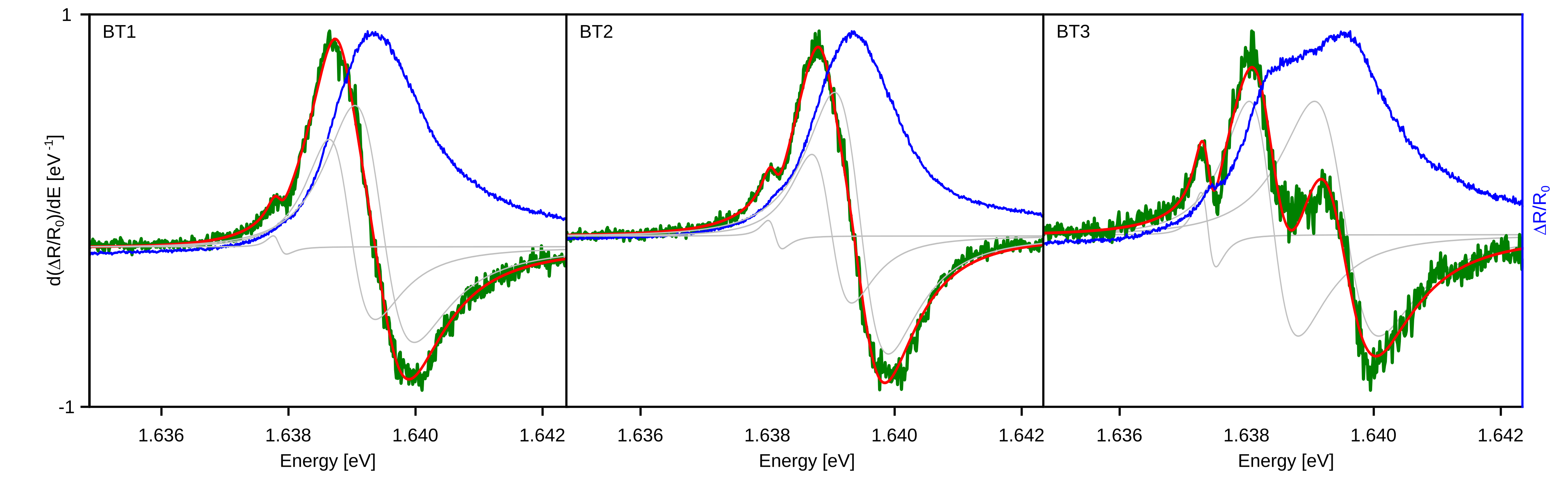}
\caption{Representative fitting results for the three different bow ties. The blue curve is the differential reflectivity spectrum $\Delta R/R_{0}$, the green one is its energy derivative, $d(\Delta R/R_{0}$)/dE, and the red one is the triple dispersive Lorentzian fit. The thin grey lines represent each individual derivated Lorentzian used to fit the derivative differential reflectivity.}
\label{FitFig}
\end{center}
\end{figure*}

The spectra present different features: the trapped exciton signal and a broad resonance for the 2D exciton. Three complex Lorentzian are used to fit the spectra, using the same phase parameter $\theta$. Two Lorentzians are needed to fit the 2D exciton, most likely due to local strain in the TMD layer. Fig.\,\ref{FitFig} present an example of fit for the three different bow ties.

\subsection{Fabrication method}
\label{Fabrication method}
All hBN, graphite, and MoSe$_2$ flakes used in this work are produced with mechanical exfoliation using scotch tape. The high-quality hBN crystal is from NIMS, and the MoSe$_2$ crystal is from HQ graphene. The vdW heterostructure is assembled using the standard PC (polycarbonate) dry transfer method in an ambient environment. The heterostructure is then dropped down to a SiO$_2$/Si substrate at 180\,C. Multilayer graphite flakes are picked up as contacts to the MoSe$_2$ monolayer. 

Nanometer-scale features are generated with e-beam lithography (Raith Voyager, 50\,keV) using a thin PMMA layer of 100 nm thick. For the quantum ring samples, the vdW heterostructure consists of hBN capping layer/top Gr gate/top hBN/MoSe$_2$ monolayer/bottom hBN/bottom Gr gate. After e-beam patterning, the hBN capping layer and the graphite top gate are etched by reactive ion etching (Oxford Plasma Pro 80) sequentially.  10\,sccm CHF$_3$/10\,sccm Ar/2\,sccm O$_2$ is used to etch hBN, and 10\,sccm O$_2$ is used to etch graphite. For the bow tie samples, the heterostructure consists of top hBN/MoSe$_2$ monolayer/bottom hBN/bottom Gr gate. The same e-beam lithography process is used to pattern the nanogap, then a thin layer of metal (3\,nm Ti/10\,nm Au) is deposited as the top gates with a Kurt J. Lesker high vacuum e-beam evaporator.

The leads and bonding pads to the patterned gates and the sample contacts are formed by 5\,nm Ti/45\,nm Au, with patterns generated by either e-beam lithography (with 200\,nm PMMA) or photo-lithography (ML3 MicroWriter).

The thicknesses of the top and bottom hBN of the devices discussed in the main text is listed in table \ref{tab:devices}.

\begin{table}[h]
\caption{ Top and bottom hBN thicknesses of the
devices discussed in the main text.}
\begin{tabular}{|c|c|c|}
\hline
Relevant figure                      & top hBN [nm] & bottom hBN [nm] \\ \hline
Fig.\,\ref{fig:two}\,\bfB,\,\bfD,\,\bfE              & 40           & 40              \\ \hline
Fig.\,\ref{fig:two}\,\bfF - \bfI, Fig.\,\ref{fig:four}\,\bfA-\bfC & 40           & 20              \\ \hline
Fig.\,\ref{fig:three}, Fig.\,\ref{fig:four}\,\bfD, \bfE            & 15           & 25              \\ \hline
\end{tabular}
\label{tab:devices}
\end{table}

\subsection{Electrostatic simulation method}
\label{Simulation method}
The electrostatic simulation is done with COMSOL multiphysics electrostatic modeling on a 3D grid using finite element method. The TMD monolayer is modeled as a semiconductor, sandwiched in two insulating hBN slabs. The bottom gate is global and the top gate is set to be certain geometry. See ref.\cite{Thureja2022} for detailed parameters.

After the trapping potential is obtained from COMSOL simulation, the confined state energy and wavefunction is obtained by solving the 2D Schrodinger equation numerically in the c.o.m frame. 

\subsection{Doping diagram}
\label{Doping diagram}

\begin{figure}[h]
\begin{center}
\includegraphics[width=\linewidth]{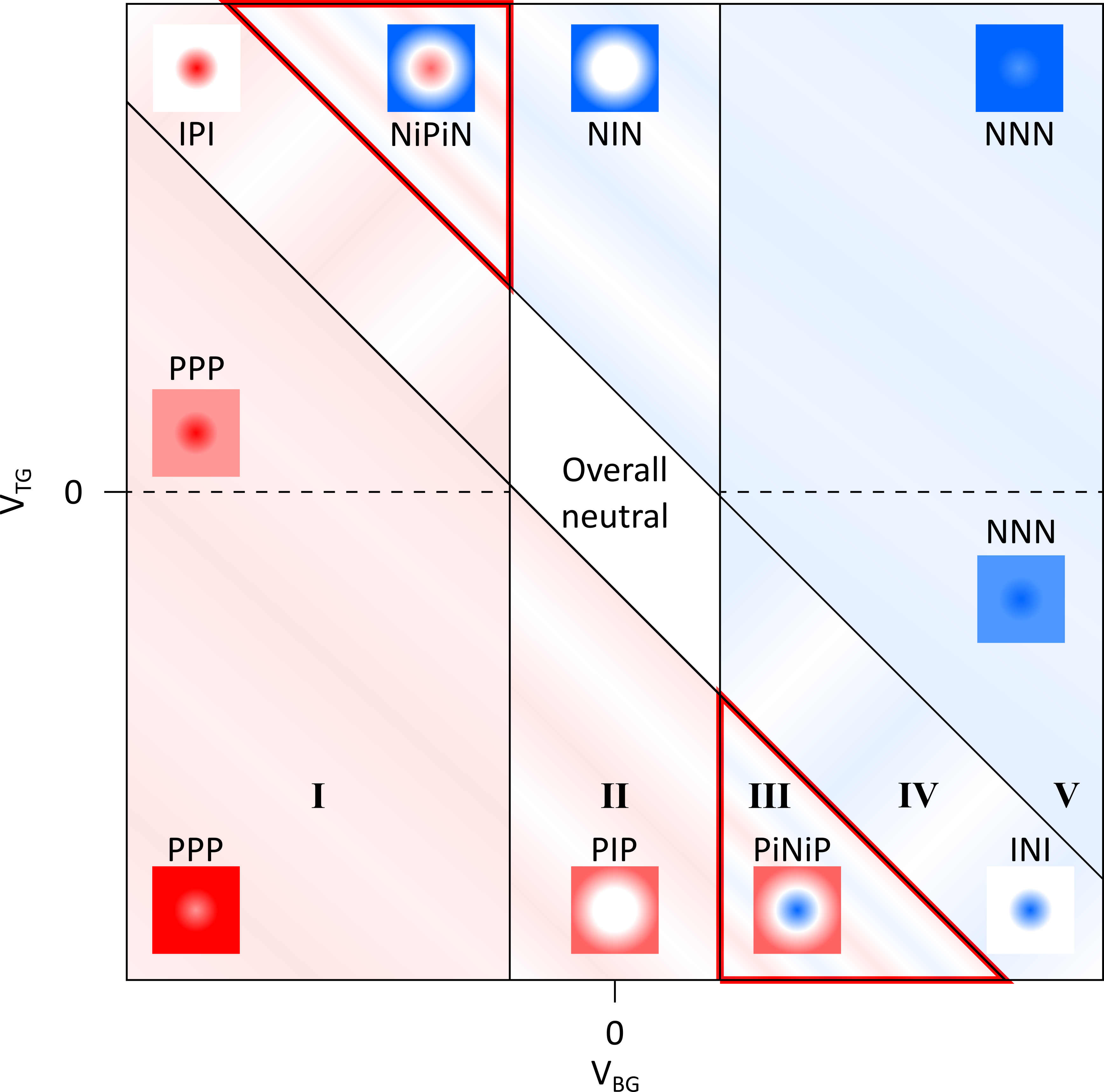}
\caption{Doping diagram for a dual gated TMD with a hole in the top gate. Each region separated by a solid black line correspond to a different doping state of the TMD. This doping state is sketched in the squares, that are charge density map, from hole doped in red to electron doped in blue, passing by white at neutrality. The thick red triangles highlight the PiNiP and NiPiN region, where the TMD 2D exciton can be trapped into a ring. The bold Roman numerals correspond to the doping regime in Fig.\,\ref{fig:two} of the main text. The two red triangles correspond to the region where ring confinement occurs.}
\label{FigDopingDiagram}
\end{center}
\end{figure}

Fig.\,\ref{FigDopingDiagram} presents a sketch of the top and bottom gate voltage dependence of the doping for a nanohole structure. First, the sample can be separated into two different region, the dual gated region outside the hole (region B in the main text), and the single gated region inside the hole (region A in the main text).

In region A, only the bottom gate acts to set the charge density of the TMD. Therefore, within the hole, the doping will switch from p- to n-doped due to the sole bottom gate. This result in the vertical lines on the doping diagram (Fig.\,\ref{FigDopingDiagram}), which are the bottom gate voltage at which the TMD switch from neutral to p- or n-doped.

The dual gated region however, one can tune separately the vertical electric field and the TMD doping. An opposite voltage between the top and bottom gate will result in an out of plane electric field and no doping, which is the central diagonal line of Fig.\,\ref{FigDopingDiagram}. On each side of this diagonal, p- and n-doped region of the outter hole appears.

Since the trapping of 1D edge states occurs in the p-i-n regions, the excitonic ring trap only appear in the upper left and bottom right red triangles (Fig.\,\ref{FigDopingDiagram}), where the doping in the hole is of a different kind than the doping outside it.

In addition, a weaker trap can be present in theory for the repulsive polaron, when both top gate and bottom gate are positive or negative. In this case, the inner part of the hole would be less doped, which will results in a lower energy polaron in the hole, and a higher one outside. The dashed line in Fig.\,\ref{FigDopingDiagram} shows the limit at which the top gate voltage changes it sign which will induce a lower charge density outside of the hole and a larger one within. This latter configuration lead to an anti-trap for the repulsive polaron. 

\subsection{Trapping potential vs $\vbg$ in the 600 nm hole}
\label{Simulation600nmhole}

\begin{figure}[h]
\begin{center}
\includegraphics[width=\linewidth]{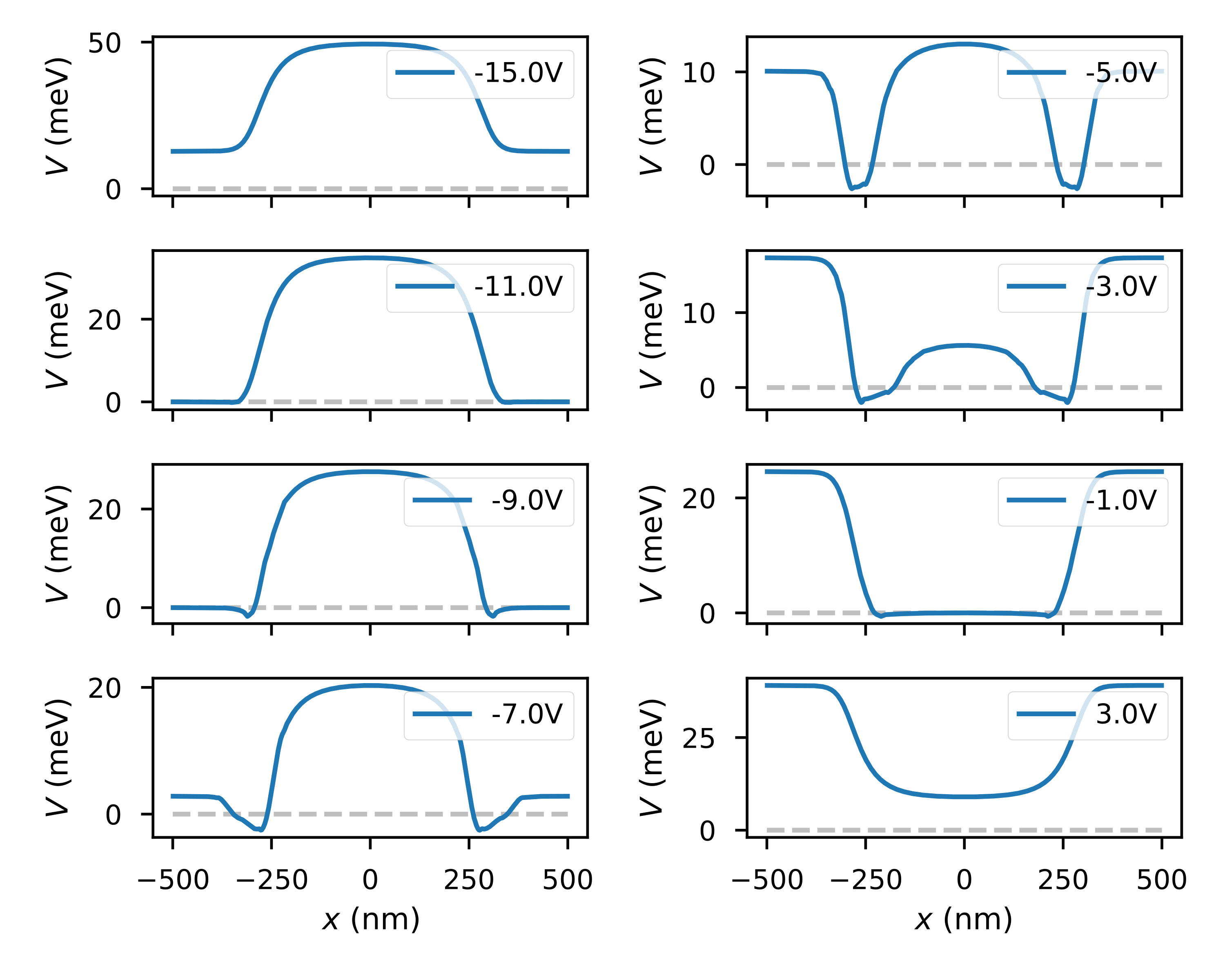}
\caption{Electrostatic simulation of the trapping potential at different $\vbg$ in the 600\,nm hole. $\vtg$ is set to be $9.5\,$V, the same as in Fig.\,\ref{fig:two}\bfE.}
\label{FigS4}
\end{center}
\end{figure}

Fig.\,\ref{FigS4} shows the simulated trapping potential at different $\vbg$ in a 600 nm hole. Here $\vtg$ is set to be $9.5\,$V, the same as in Fig.\,\ref{fig:two}\,\bfE. A ring-shaped trap is only formed when $\vbg$ is from $-9\,$V to $-1\,$V, which agrees with our observation in Fig.\,\ref{fig:two}\,\bfE~that the trapped states only show up in regime III. 

\subsection{Simulation results for tunable 0D states in the bow tie}
\label{SimulationBT0D}
Electrostatic simulation for a nano-gap structure made with thin metal layers can be tricky. Due to the small size of the electrode and reduced conductivity, the voltage applied from the source-measurement unit might not reflect the actual voltage drop across the bow tie. This results in a much larger energy dispersion in the simulation compared to the measurements. 

\begin{figure}[htbp]
\begin{center}
\includegraphics[width=\linewidth]{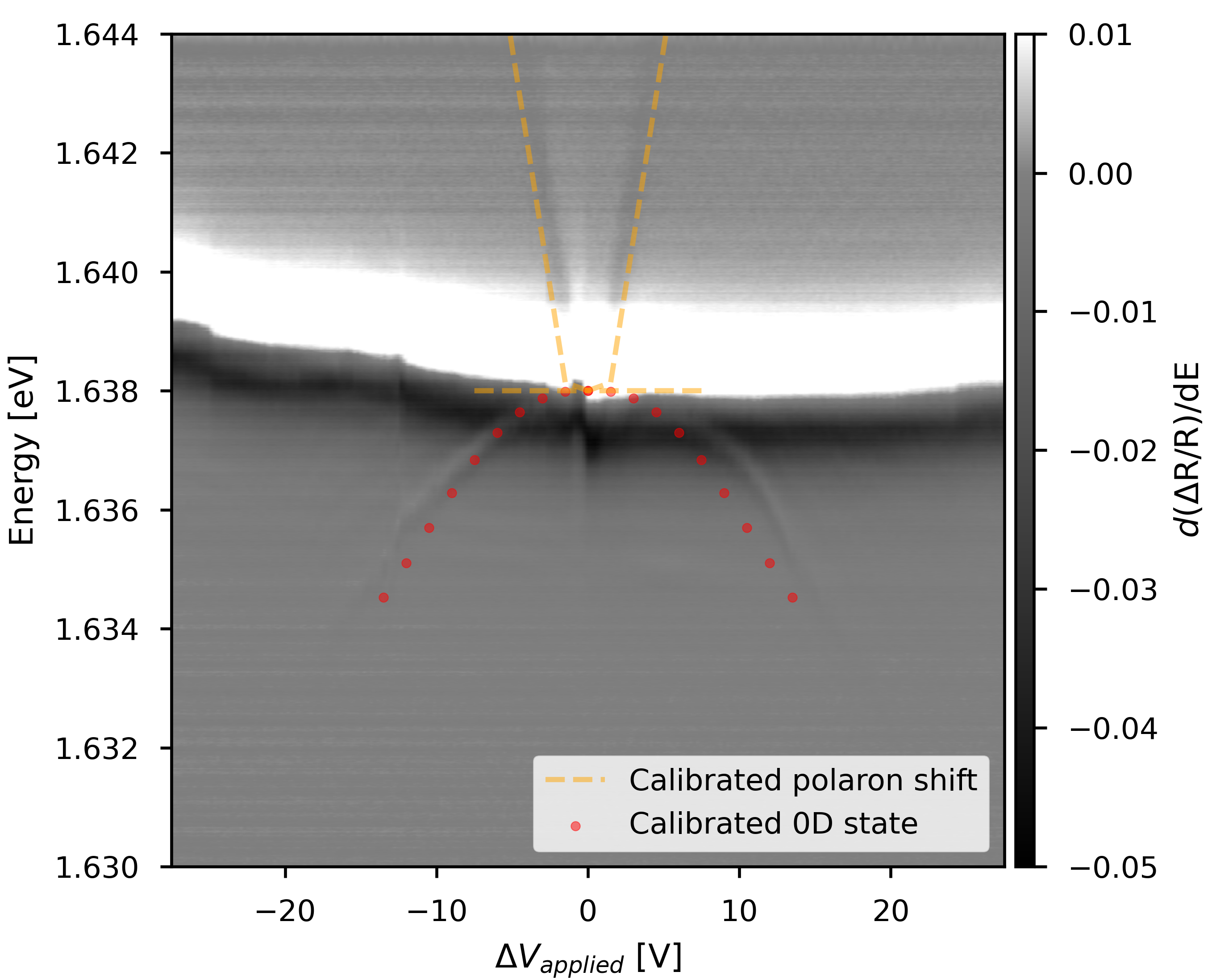}
\caption{Comparison of the calculated 0D trapped state dispersion and the measurement. The $\Delta V$ for the simulation input is calibrated using the blue shifted repulsive polaron state. }
\label{FigBTSimulation}
\end{center}
\end{figure}

Here we present a way to calibrate the actual $\Delta V$ with respect to the applied $\Delta V$. In the reflectivity spectra in Fig.\,\ref{fig:three}\,\bfF, we observe both the 0D states and the blue-shifting repulsive polaron (RP) emerging from charging the TMD below the left and right electrodes. The blueshift of the RP corresponds to the second term in Eq.\ref{eq:TotalPotential}, which can be calculated from the simulated charge density. By comparing it to the measured RP blueshift, we determine a reduction factor that overlaps the simulated polaron shift to the measured polaron shift, as shown in Fig.\,\ref{FigBTSimulation} (here $\Delta V_{\text{actual}} = 0.33 \Delta V_{\text{applied}}$). Then we use the same reduction factor to calculate the trapping potential and solve the energy dispersion of the 0D states (shown in red dots in Fig.\,\ref{FigBTSimulation}). It agrees reasonably well with the measurement. 

\subsection{Transition from 0D confinement to 1D confinement in the bow tie}
\label{BT0Dand1D}

\begin{figure*}[htbp]
\begin{center}
\includegraphics[width=0.8\linewidth]{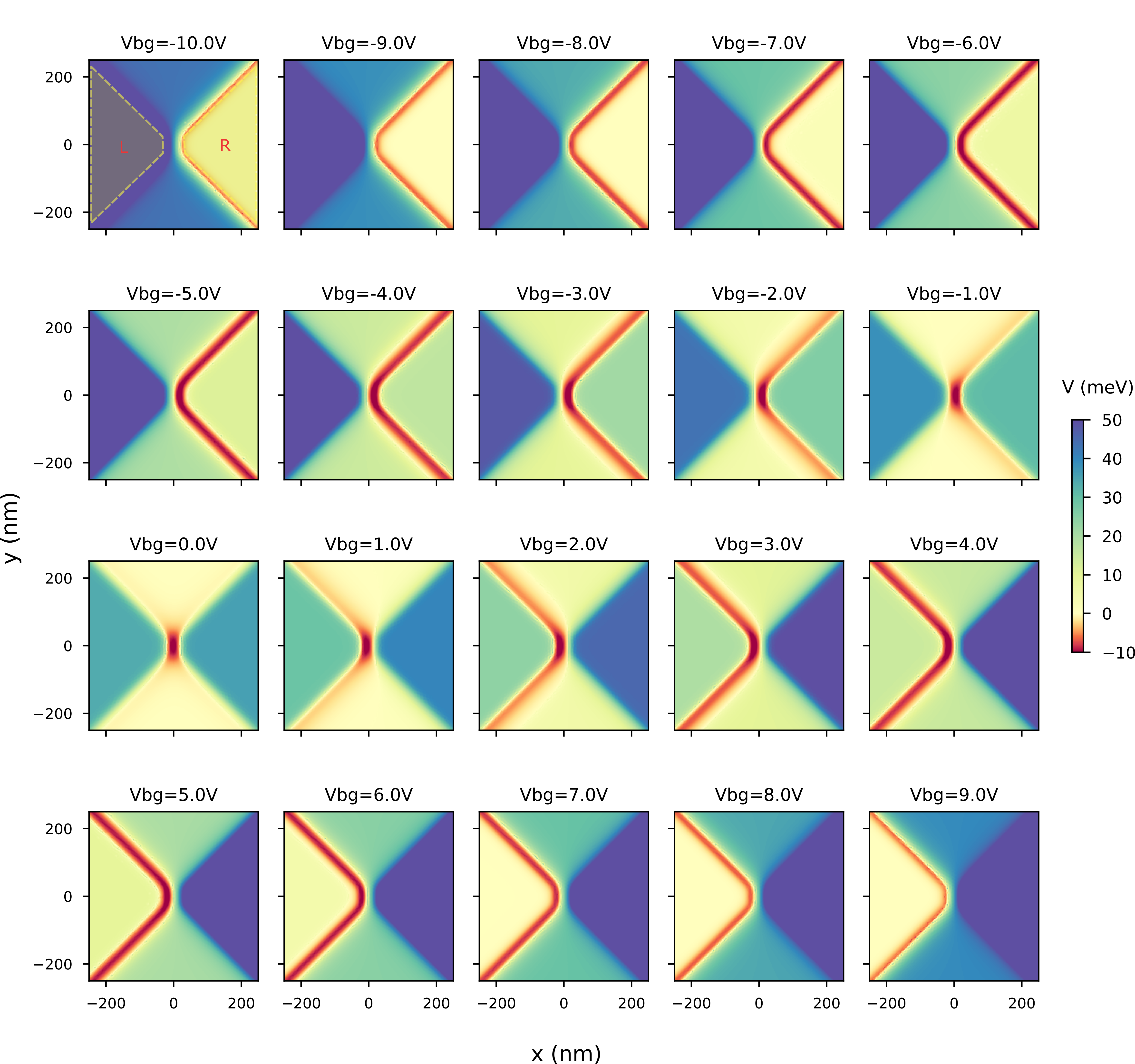}
\caption{Simulation results for the evolution of the potential landscape with the backgate voltage at $(\vL, \vR)=(5\,\text{V},-5\,\text{V})$. It shows a continuous transition from the 0D dot confinement ($\vbg=0\,$V) to 1D edge confinement wrapped around the right (left) electrode when the backgate is tuned to negative (positive) voltage. }
\label{ExcitonBS_sim}
\end{center}
\end{figure*}

\begin{figure*}[htbp]
\begin{center}
\includegraphics[width=0.4\linewidth]{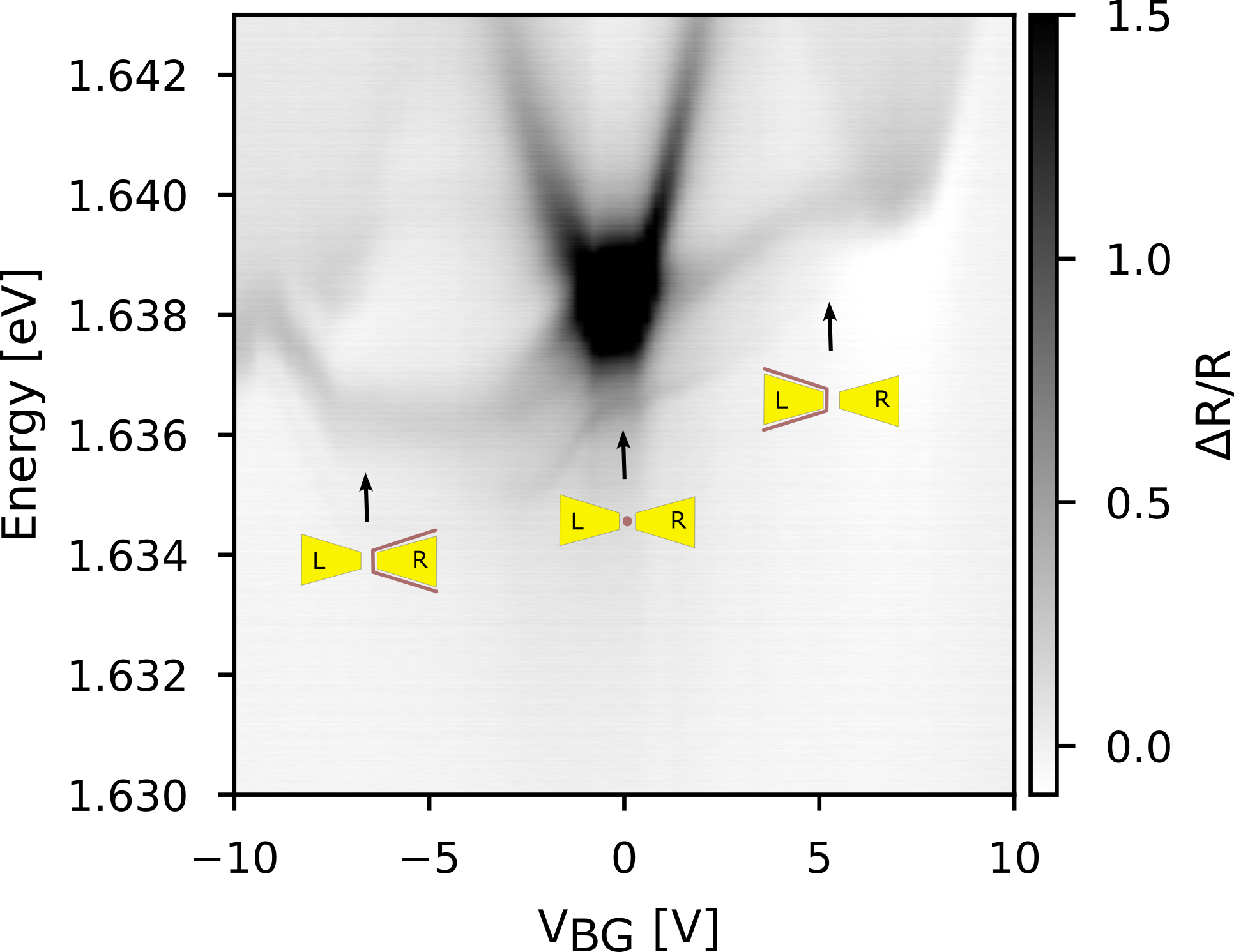}
\caption{Reflectivity spectra in the bow tie structure as a function of $\vbg$ at $(\vL, \vR)=(5\,\text{V},-5\,\text{V})$. }
\label{FigBGscan}
\end{center}
\end{figure*}

Here in Fig.\,\ref{ExcitonBS_sim}, we show the simulation results for the potential landscape in the bow tie structure, as a function of the backgate voltage. The left and right bow tie electrodes are at +5\,V and -5\,V respectively, i.e. $\Delta V = 10\,$V. It demonstrates a continuous transition from the 0D dot confinement ($\vbg=0\,$V) to 1D edge confinement wrapped around the right (left) electrode when $\vbg$ is tuned to negative (positive) values. This 0D to 1D transition is observed experimentally as well, see Fig.\,\ref{FigBGscan}.  

\begin{figure*}[htbp]
\begin{center}
\includegraphics[width=0.8\linewidth]{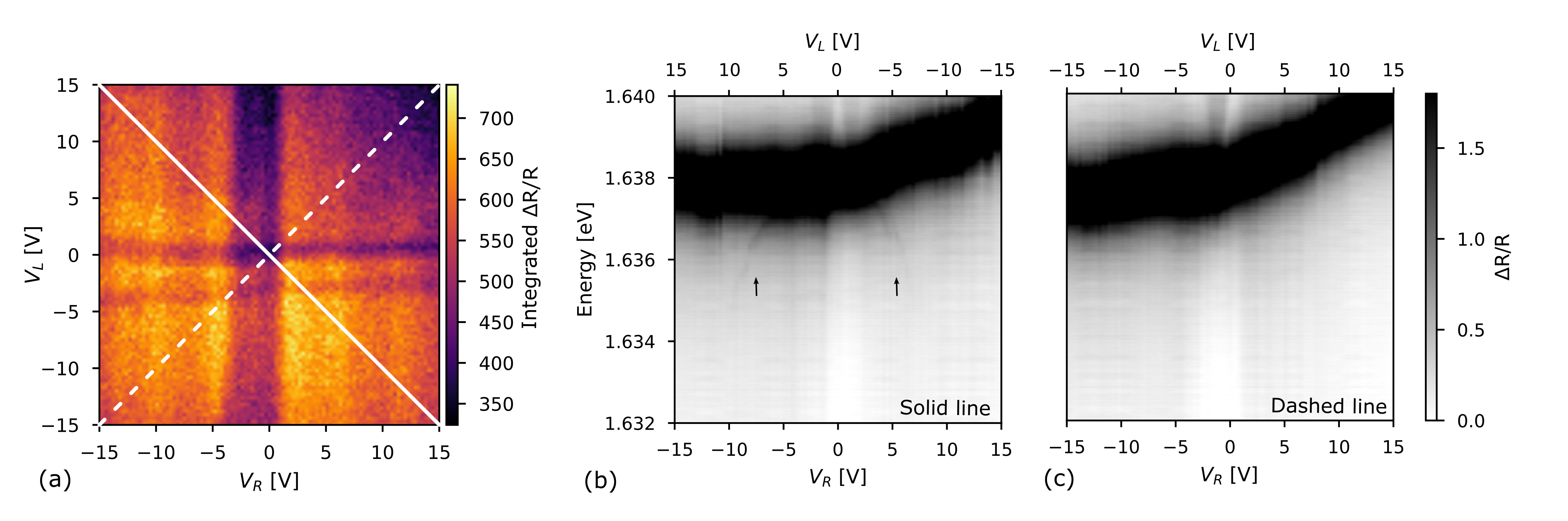}
\caption{$\vL-\vR$ dual gate scan map of reflectivity at $\vbg=0\,$V. (\bfA) The integrated reflection map, the vertical (horizontal) stripe shows the gate range where the 2D exciton under the right (left) bow tie electrode is neutralized. (\bfB) The $\Delta R/R$ line cut along the solid line in (\bfA) which corresponds to applying $\Delta V$ across the bow tie. As expected, signature for 0D trapped state is present. (\bfC) The $\Delta R/R$ line cut along the dashed line in (\bfA) corresponding to $\Delta V = 0\,$V, where no trapped state is observed. }
\label{FigS5}
\end{center}
\end{figure*}

Fig.\,\ref{FigS5} shows a $\vL-\vR$ dual gate reflection contrast scan at $\vbg = 0\,$V. Panel \bfA~is the integrated reflection map, and the vertical (horizontal) stripe shows the gates range where the 2D exciton under the right (left) bow tie electrode is kept neutral. Panel \bfB~is the $\Delta R/R_0$ line cut along the solid line in panel \bfA~which corresponds to applying $\Delta V$ across the bow tie. As expected, signature for 0D trapped states is present. Panel \bfC~is the $\Delta R/R_0$ line cut along the dashed line in panel \bfA, corresponding to $\Delta V = 0\,$V. No 0D state is observed because no trap is formed when the same voltage is applied to both bow tie electrodes. 

\begin{figure*}[htbp]
\begin{center}
\includegraphics[width=\linewidth]{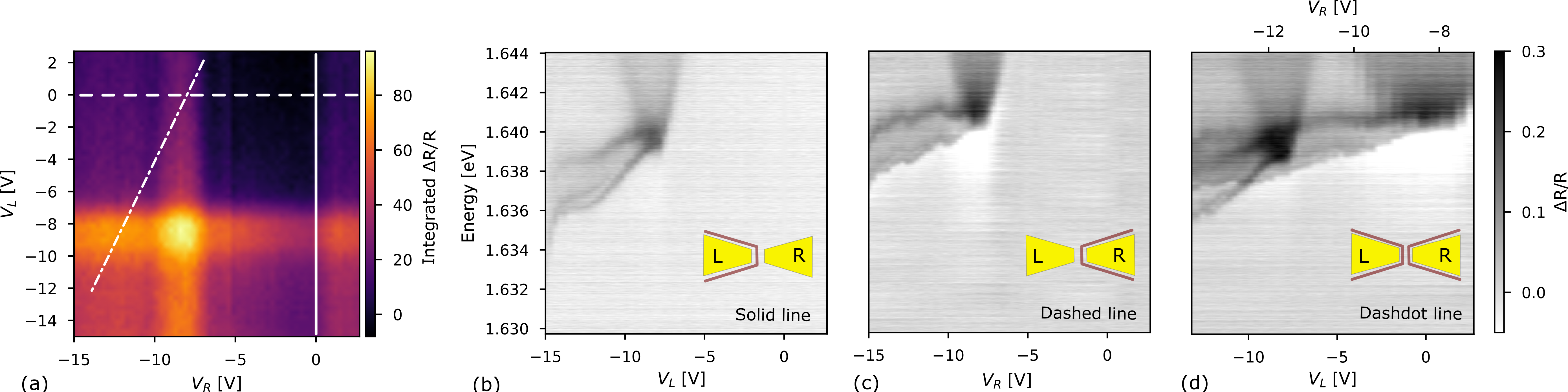}
\caption{$\vL-\vR$ dual gate scan map of reflectivity at $\vbg=10\,$V. (\bfA) The integrated reflection map, the vertical (horizontal) stripes shows the gates range where the 2D exciton under the right (left) bow tie electrode is neutralized. (\bfB) The $\Delta R/R$ line cut along the solid line at $\vR=0\,$V, displaying 1D confined states around the left bow tie electrode (as shown in the inset). (\bfC) The $\Delta R/R$ line cut along the dashed line at $\vL=0\,$V, displaying 1D confined states around the right bow tie electrode.  (\bfD) The $\Delta R/R$ line cut along the dashdot line, showing 1D confined states around both bow tie electrode. }
\label{FigS6}
\end{center}
\end{figure*}

Fig.\,\ref{FigS6}  shows a $\vL-\vR$ dual gate scan map of reflectivity at $\vbg=10\,$V.  Panel \bfA~is the integrated reflection map, and the vertical (horizontal) stripe indicates the gates range where the 2D exciton under the right (left) bow tie electrode is tuned to neutrality. This is at a negative voltage now because $\vbg$ is set to be positive. Panel \bfB~is the $\Delta R/R_0$ line cut along the solid line at $\vR=0\,$V, showing 1D confined states around the left bow tie electrode (as shown in the inset). Panel \bfC~is the $\Delta R/R_0$ line cut along the dashed line at $\vL=0\,$V, showing 1D confined states around the right bow tie electrode. Panel \bfD~plots the line cut along the dashdot line, displaying two groups of 1D confined states around both bow tie electrodes. The two sets of 1D confined states can also be tuned to resonance. This shows that bow tie structures are very versatile and can host both 0D and 1D tunable confined exciton.

\subsection{0D trapped states in multiple samples}
\label{BTsamples}
The 0D trapped state is consistently observed in various bow tie structures on multiple samples, as shown in Fig.\,\ref{FigS3}. For some bow ties, we only see 0D state on one side of $\Delta V$, possibly due to charge inhomogeneity around the nano-gap region. 

\subsection{Polarization dependence of the 0D states}
\label{BTpol}
\begin{figure*}[h]
\begin{center}
\includegraphics[width=0.5\linewidth]{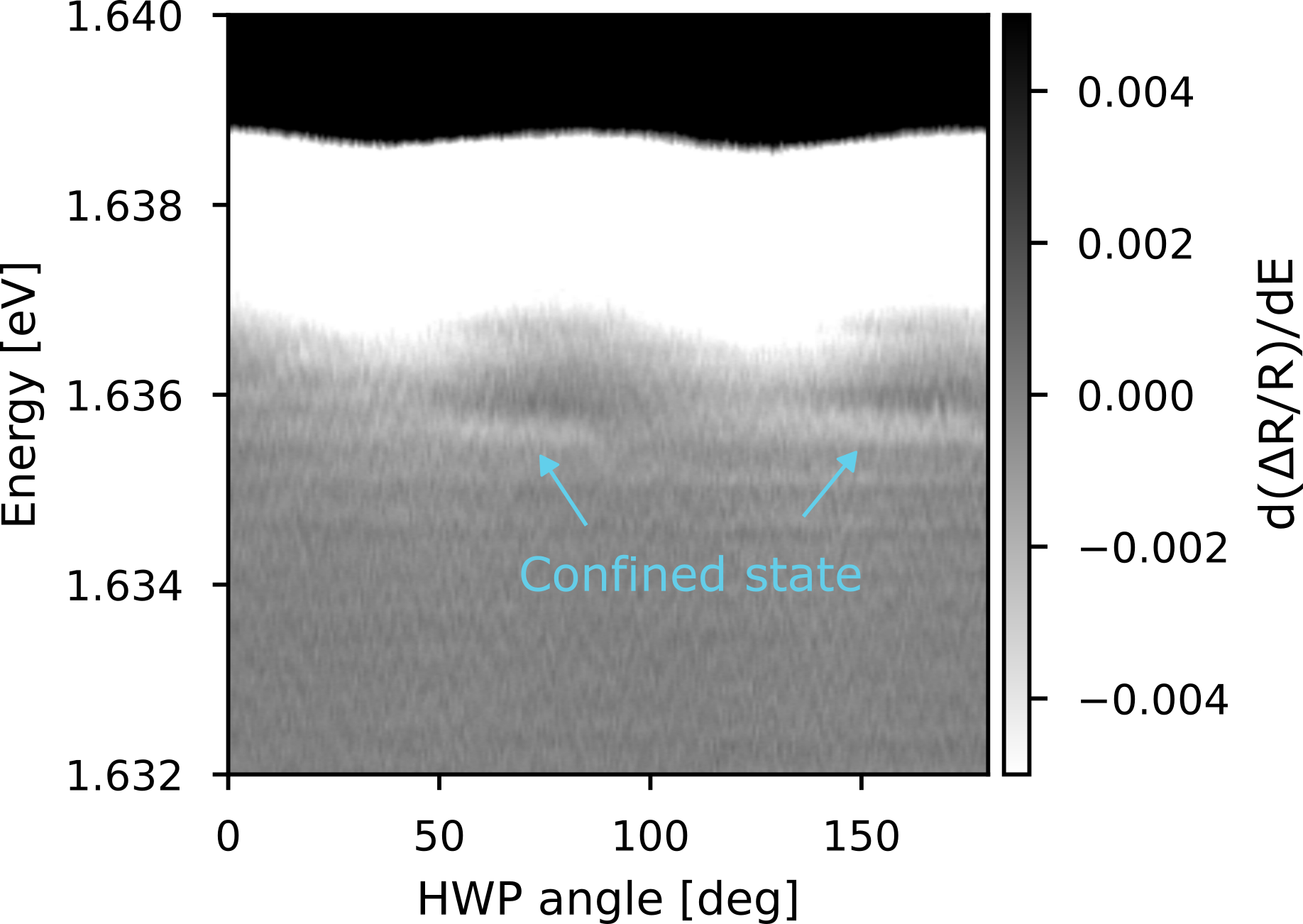}
\caption{Polarization dependence of the 0D trapped states in the bow tie structure.}
\label{FigS7}
\end{center}
\end{figure*}

\begin{figure*}[h]
\begin{center}
\includegraphics[width=\linewidth]{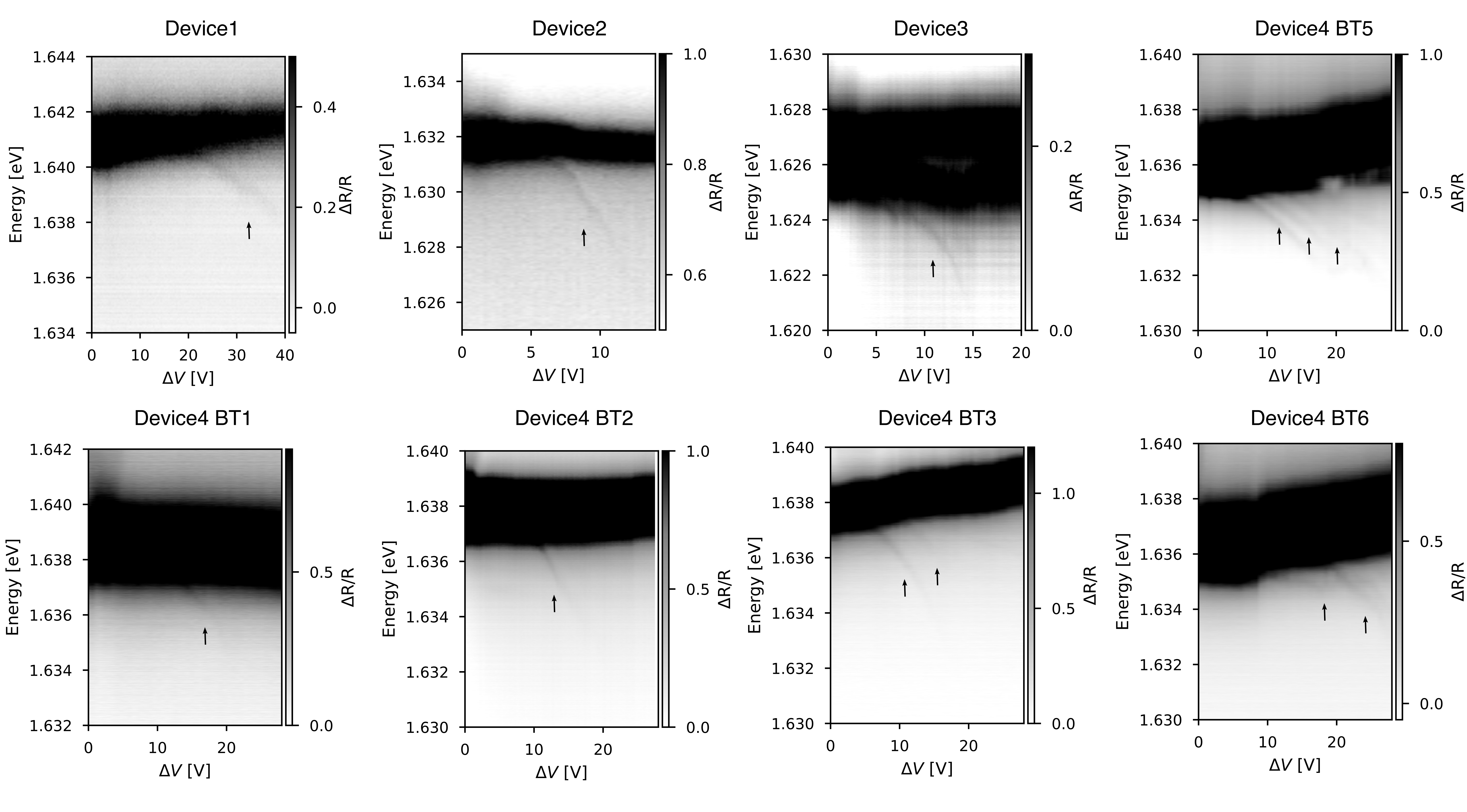}
\caption{0D confined states measured in bow tie structures on multiple samples.}
\label{FigS3}
\end{center}
\end{figure*}

In Fig.\,\ref{FigS7}, we show the polarization dependence of the 0D trapped states in the bow tie structure. The trapped state follows a linear polarization basis, consistent with the anisotropy of the trapping potential shown in Fig.\,\ref{fig:three}\,\bfD. 

\end{document}